\documentclass{article}

\usepackage{arxiv}
\usepackage{tabularx}
\usepackage[ruled,linesnumbered,commentsnumbered]{algorithm2e}
\usepackage[utf8]{inputenc}

\usepackage[bookmarks=false]{hyperref}
\usepackage{listings}
\usepackage{color}
\usepackage{graphicx}%
\usepackage[utf8]{inputenc} 
\usepackage[T1]{fontenc}    
\usepackage{url}            
\usepackage{booktabs}       
\usepackage{amsfonts}       
\usepackage{nicefrac}       
\usepackage{microtype}      
\usepackage{lipsum}
\usepackage{mathptmx}
\usepackage{hyperref}
\usepackage{graphicx}
\usepackage{subcaption} 
\usepackage{footnote}
\usepackage{amsmath}
\usepackage{amssymb}
\usepackage{mathrsfs}
\usepackage{array}
\usepackage{cite}
\usepackage{multirow}
\usepackage{siunitx}
\usepackage{tablefootnote}
\usepackage{mathtools}
\usepackage{commath}
\usepackage{scalerel}
\usepackage[utf8]{inputenc}
\title{When Noise meets Chaos: Stochastic Resonance in Neurochaos Learning}
\author{
Harikrishnan NB, Nithin Nagaraj\\
Consciousness Studies Programme,\\ National Institute of Advanced Studies,\\ Indian Institute of Science Campus, Bengaluru, India. \\  \texttt{harikrishnannb@nias.res.in, nithin@nias.res.in  } \\
}
\begin{document}
\maketitle
\begin{abstract}
Chaos and Noise are ubiquitous in the Brain. Inspired by the chaotic firing of neurons and the constructive role of noise in neuronal models, we for the first time connect chaos, noise and learning. In this paper, we demonstrate Stochastic Resonance (SR) phenomenon in Neurochaos Learning (NL). SR manifests at the level of a single neuron of NL and enables efficient subthreshold signal detection. Furthermore, SR is shown to occur in single and multiple neuronal NL architecture for classification tasks - both on simulated and real-world spoken digit datasets. Intermediate levels of noise in neurochaos learning enables peak performance in classification tasks thus highlighting the role of SR in AI applications, especially in brain inspired learning architectures.   
\end{abstract}

\section{Introduction}
The discipline of `Artificial Intelligence' (AI) originated with the aim of building computer systems that mimics the human brain. This involves the interplay of neuroscience and computational/mathematical models. Over the years since the inception of AI, both neuroscience and computational approaches have expanded their boundaries. This in turn shifted the focus of AI from building systems by exploiting the properties of brain to mere engineering point of view i.e., `what works is ultimately all that really matters'~\cite{hassabis2017neuroscience_AI}. The engineering approaches like optimization and hyperparameter tuning evaluate AI from a performance point of view. This particular view greatly limits the original motivation of AI. In this research, we use two key ideas from neuroscience namely Chaos and Stochastic Resonance to develop novel machine learning algorithms.

With the current understanding, there are nearly 86 billion neurons~\cite{azevedo2009equal_neurons_in_brain} in the human brain. They interact with each other to form a complex network of neurons. 
These neurons are inherently non-linear and found to exhibit a fluctuating neural response for the same stimuli on different trails while doing experiments. The fluctuating neural response is in part due to (a) \emph{inherent chaotic nature of neurons}~\cite{therechaos2}. Chaotic neurons are sensitive to initial states and thus show fluctuating behaviour to varying initial neural activity with the start of each trail. The second source of fluctuating behaviour can be attributed to (b) \emph{neuronal noise and interference}~\cite{faisal2008noise_nervous_system}.
Noise has its effect on the perception of sensory signals to the motor response generation~\cite{faisal2008noise_nervous_system}. Thus, noise poses a challenge as well as a benefit to information processing. The research in noise can be traced back to the experiment of Robert Brown in 1822~\cite{brown1828brief}. In the experiment, the Scottish botanist observed under a microscope the irregular movement of pollen on the surface of a film of water. Robert Brown tried to investigate the reason behind this irregular fluctuations. This phenomenon is known as Brownian motion. This problem was successfully solved by Albert Einstein in 1905~\cite{einstein1905brownian}. The noise produced by the Brownian motion is termed as brown noise or red noise. The term brown is indicated to give credit to Robert Brown for his key observations and laying out experiments to understand Brownian motion. Another interesting research in 1912 by Dutch Physicist and the first woman in noise theory, Geertruida de Haas-Lorentz viewed electrons as Brownian particles. This inspired the Swedish Physicist Gustav Adolf Ising in 1926 to explain why galvanometers cannot be cascaded indefinitely to increase amplification~\cite{ising1926lgalvanometer}. The next leap in noise research was brought by J. B. Johnson and H. Nyquist. During the year 1927-1928 Johnson published his well known thermal voltage noise formula and derived the formula theoretically in collaboration with Nyquist~\cite{nyquist1928thermal, johnson1928thermal}. The world took a turn in 1948 by the ground breaking work of Claude Elwood Shannon who created the field called Information theory~\cite{shannon1948mathematical}. In his 1948 paper titled ``A Mathematical Theory of Communication", Shannon solved how to reliably transmit a message through an unreliable (noisy) channel. Shannon showed that any communication channel can be modeled in terms of bandwidth and noise. Bandwidth is the range of electromagnetic frequencies required to transmit a signal and noise is an unwanted signal that disrupts the original signal. He further showed how to calculate the maximum rate at which a data can be sent through a channel with a particular bandwidth and noise characteristics with zero error. This is called the rate of channel capacity or Shannon limit~\cite{shannon1948mathematical}.

    All these research, especially Shannon's work, considered noise as an unwanted signal that adversely affects the communication. But in the second half of 20th century, the constructive advantage of noise in signal detection and also the advantages of noise in physiological experiments lead to the birth of \emph{Stochastic Resonance}. The term Stochastic Resonance (SR) was first used in the context of noise optimized systems by Roberto Benzi~\cite{berger2012climatic} in 1980 with regard to a discussion on climate variations and variability. Benzi introduced SR in connection with the explanation to a periodicity of $10^{5}$ years found in the power spectrum of
paleoclimatic variations for the last $700,000$ years~\cite{benzi1982_climate_SR}. The energy balance models failed to explain this phenomenon. Benzi, in his paper titled ``Stochastic Resonance in Climate Change"~\cite{benzi1982_climate_SR}, suggested that the combination of internal stochastic perturbations along with external periodic forcing due to earth's orbital variations are the reasons behind the ice age cycle. The paper further suggests that neither the stochastic perturbations nor periodic forcing alone can reproduce the strong peak found at a periodicity of $10^5$ years. Thus, the effect produced by this co-operation of noise and periodic forcing was termed as Stochastic Resonance by Benzi. Even though this explanation is still a subject of debate, but the definition of the term SR got evolved in the coming years. SR finds application in climate modelling~\cite{benzi1982_climate_SR}, electronic circuits~\cite{fauve1983SR_electronic_circuit}, neural models~\cite{bulsara1991_SR_neuron}, chemical reactions~\cite{leonard1994_SR_chemical_reaction}, etc. The original use of the resonance part of SR comes from the plot of output signal to noise ratio (SNR) that exhibits a single maximum for an intermediate intensity of noise ~\cite{mcdonnell2008stochastic}. A motivating idea to develop SR based electronic devices or neuromorphic systems comes from the brain, because we know that the brain is far better than electronic devices in terms of low computational power, robust to noise and neural interference. SR in the brain and nervous system could serve as a motivation for the design of robust machine learning architectures and electronic systems.
The observation of SR in neural models was first published in 1991~\cite{bulsara1991_SR_neuron}. The research in SR accelerated when the 1993 Nature article reported the presence of SR in physiological experiments on crayfish mechanorecetors~\cite{douglass1993_SR_cray_fish}. In the same year, yet another highly cited paper -- SR on neuronal model~\cite{longtin1993_SR_neuron_model} -- became widely popular. These research studies triggered the expansion of SR  in mathematical models of neurons, biological experiments, behavioural experiments especially in paddle fish~\cite{russell1999paddle_fish}, noise enhanced cochlear implants~\cite{chatterjee2001noise_cochlear_implant} and computations~\cite{bulsara2010logical}. Despite the wide popularity of SR, only a few papers have focused on the application of SR in machine learning (ML) and deep learning (DL)~\cite{ikemoto2018noise_SR_NN, schilling2020intrinsic_SR_Speech}. The current ML and  DL algorithms assume ideal working conditions, i.e the input data is noiseless. But in practice, there is unavoidable noise that distorts measurements by sensors. Hence, there is a research gap from a theoretical and an implementation point of view as far as ML/DL algorithms are concerned. We define SR as noise enhanced signal processing~\cite{mcdonnell2008stochastic}. In a nutshell for SR to occur, the following four elements are required:
\begin{enumerate}
    \item An information carrying signal.
    \item A noise added to the input signal.
    \item A non-linear system.
    \item A performance measure which captures the  relationship between the output and input with respect to varying noise intensity. For classification tasks, we shall use F1-score as a measure to capture the performance of the non-linear system with respect to varying noise intensities.
 \end{enumerate}   
 In this work, we highlight how SR is inherently used in the recently proposed Neurochaos Learning (NL)~\cite{harikrishnan2020neurochaos} architecture namely \verb+ChaosNet+~\cite{balakrishnan2019chaosnet}. The sections in the paper are arranged as follows: Section~\ref{Sec_SR} describes NL and how SR is naturally manifested in NL. Section~\ref{Sec:Exp_Evidence} highlights the empirical evidence of SR in NL for both simulated and real world datasets. Section~\ref{Sec:Conclusion} deals with conclusion and future work.
\section{Stochastic Resonance in Neurochaos Learning (NL)~\label{Sec_SR}}

\begin{figure}[!h]
    \centerline{ \includegraphics[width=0.6\textwidth]{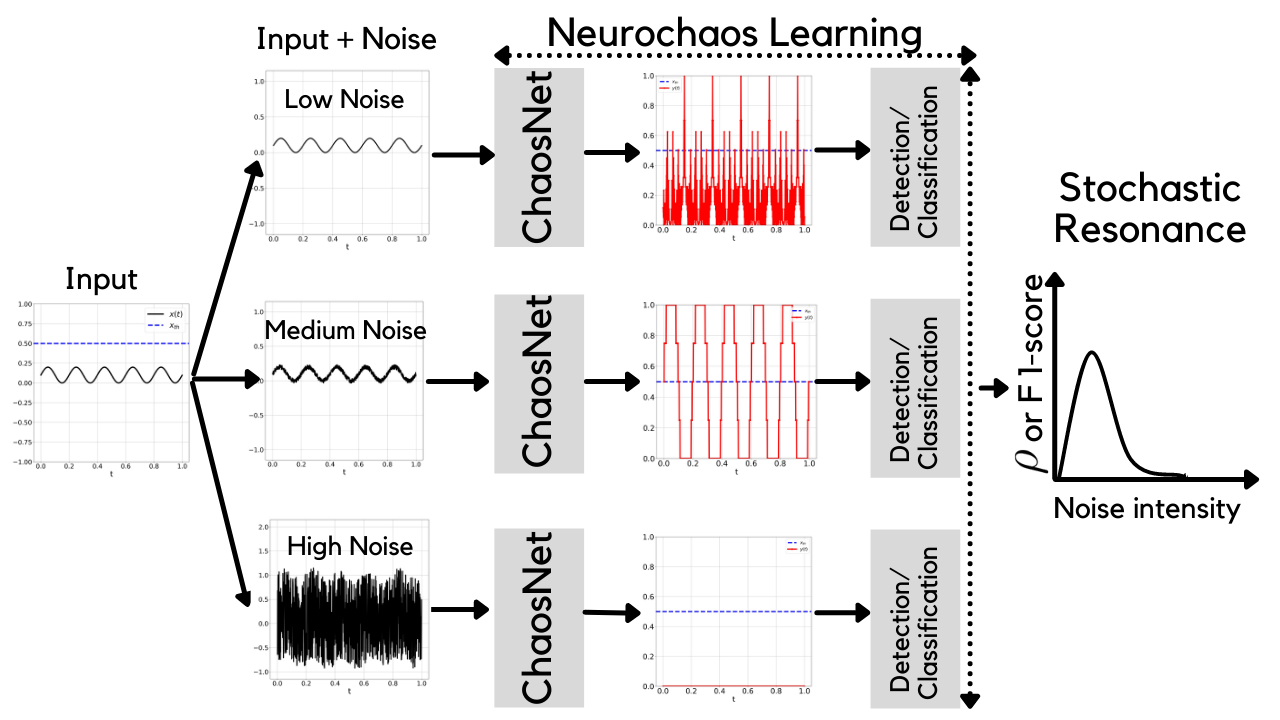}}
    
     \caption{Neurochaos Learning exhibits Stochastic Resonance for signal detection and classification. 
     }
    \label{Fig_complete_archi}
    \end{figure}

Inspired by the presence of neural chaos (neurochaos) in the brain~\cite{therechaos1, therechaos2}, we have recently proposed two machine learning  architectures namely \verb+ChaosNet+~\cite{balakrishnan2019chaosnet} and Neurochaos-SVM~\cite{harikrishnan2020neurochaos} which we term as Neurochaos Learning or NL. The inspiration of the term Neurochaos~\cite{therechaos2} comes from the chaotic behaviour exhibited at different spatiotemporal scales in the brain~\cite{therechaos1, therechaos2}. 
In this paper, we uncover the inherent constructive use of noise in NL. The crux of this paper can be represented by the flow diagram in  Figure~\ref{Fig_complete_archi}. A noisy signal is input to NL (\verb+ChaosNet+) for the purposes of either signal detection (subthreshold) or classification (from multiple classes). In the diagram, we depict three possibilities -- low, medium and high noise. As it turns out, the performance of NL for both signal detection and classification is found to be optimum for an intermediate or medium noise intensity. The performance of NL is quantified using  cross correlation coefficient ($\rho$) for signal detection and macro F1-score for classification. Thus, we claim that SR is inherent in NL and is responsible for its superior performance. The rest of this paper will showcase this idea using evidence from experiments conducted using both simulated as well as real-world examples.

\subsection{\label{sec:ChaosNet} ChaosNet: A NL architecture} 
\verb+ChaosNet+, a NL architecture, is distinctly different when compared with other machine learning algorithms.  It consists of a layer of chaotic neurons which is the 1D GLS (Generalized L\"{u}roth Series) map, $C_{GLS}: [0,1) \rightarrow [0,1)$, defined as follows: 
\begin{eqnarray*}
C_{GLS}(x)  =  \left\{\begin{matrix}
\frac{x}{b}&, ~~~~ 0 \leq x < b, \\ 
\frac{(1-x)}{(1 - b)}&, ~~~~ b \leq x < 1, 
\end{matrix}\right. \\
\end{eqnarray*}
where $x \in [0,1)$ and $0 < b <1$. 

\begin{figure}[!h]
    \centerline{ \includegraphics[width=0.45\textwidth]{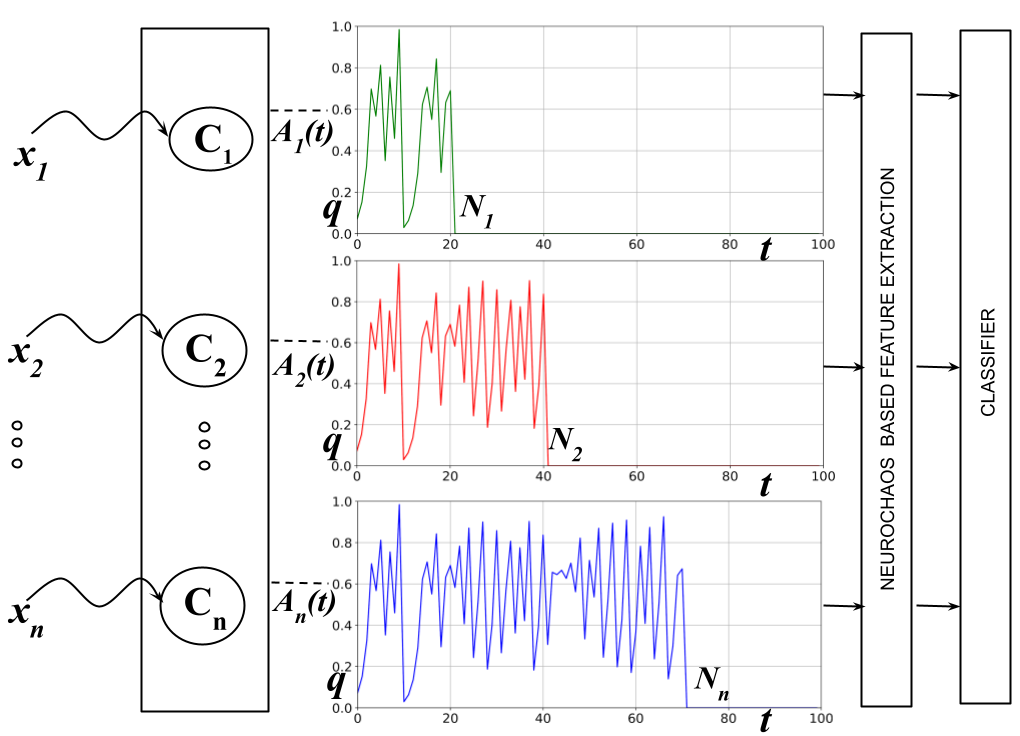}}
    
     \caption{Neurochaos Learning (NL) Architecture: 
     ChaosNet~\cite{balakrishnan2019chaosnet} is an instance of NL architecture. The input layer consists of GLS neurons. When the $r$-th stimulus (normalized) $x_r$ is input to the $r$-th GLS neuron $C_r$, it starts firing chaotically giving a neural trace. The firing stops when this neural trace matches the stimulus. Features are extracted out of the neural trace which are then passed on to a classifier.}
    \label{FIg_model_archi}
    \end{figure}

Figure~\ref{FIg_model_archi} depicts the \verb+ChaosNet+ (NL) architecture. Each GLS neuron starts firing upon encountering a stimulus (normalized input data sample) and halts when the neural firing matches the stimulus. From the neural trace of each GLS neuron, we extract the following features - \emph{firing time}, \emph{firing rate}, \emph{energy of the chaotic trajectory}, and \emph{entropy of the symbolic sequence of chaotic trajectory}~\cite{harikrishnan2020neurochaos}. \verb+ChaosNet+ architecture provided in Figure~\ref{FIg_model_archi} uses a very simple decision function - cosine similarity of mean representation vectors of each class with the chaos based feature extracted test data instances~\cite{balakrishnan2019chaosnet}. The classifier provided in Figure~\ref{FIg_model_archi} need not be restricted to SVM or cosine similarity. NL features can be combined with the state of the art DL methods. Refer ~\cite{balakrishnan2019chaosnet, harikrishnan2020neurochaos} for a detailed explanation regarding the working of the method. It is important to note that \verb+ChaosNet+ does not make use of the celebrated backpropagation algorithm (which conventional DL uses), and yet yields state-of-the-art performance for classification~\cite{balakrishnan2019chaosnet}. Very recently, \verb+ChaosNet+ has been successfully extended to fit in continual learning framework~\cite{laleh2020chaotic}.

\subsection{Stochastic resonance in a single GLS neuron}
\begin{figure}[!h]
    \centerline{ \includegraphics[width=0.45\textwidth]{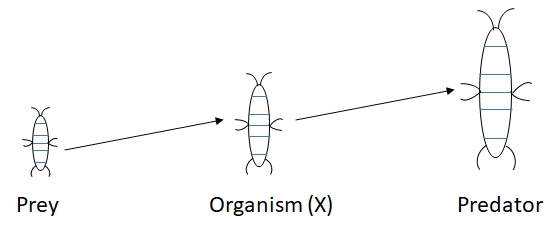}}
         \caption{Food chain of a (hypothetical) cannibalistic species. The organism $X$ feeds on other organisms of its species (prey) or is fed by other organisms of its species (predator) depending on their size. The organism $X$ has to decide whether the approaching one is food (prey) or death (predator) -- a binary classification problem. It has one internal GLS neuron which fires upon receiving stimulus which is light reflected of from the approaching organism (either predator or prey) that encodes the size of the organism. }
    \label{Fig_PP}
    \end{figure}

\begin{figure}[!h]
\centering
	\begin{subfigure}{0.45\linewidth}
		\centering
		\includegraphics[width=1\linewidth]{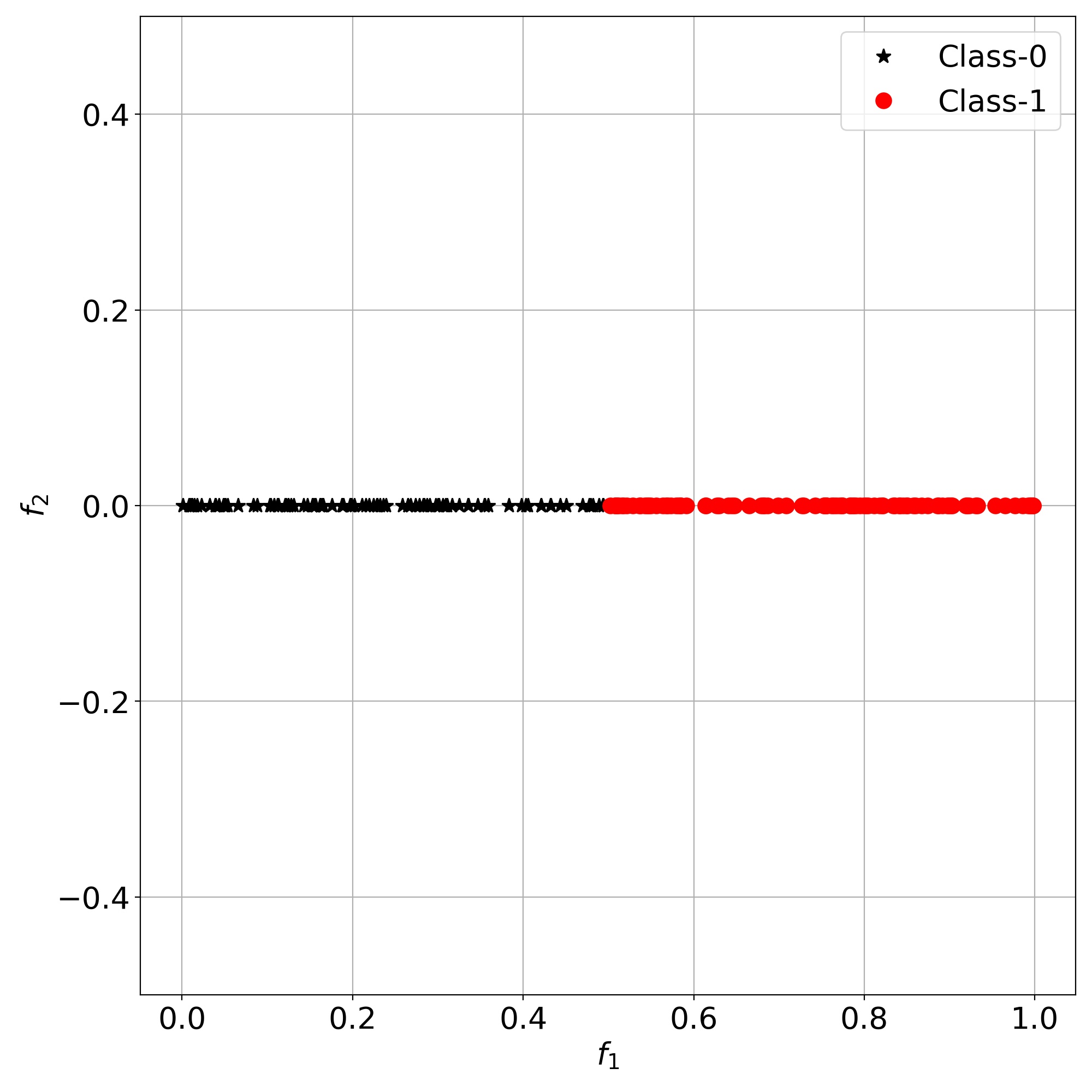}
		\caption{}\label{Fig_single_variable_Classification}
	\end{subfigure}
	\begin{subfigure}{0.45\linewidth}
		\centering
		\includegraphics[width=1\linewidth]{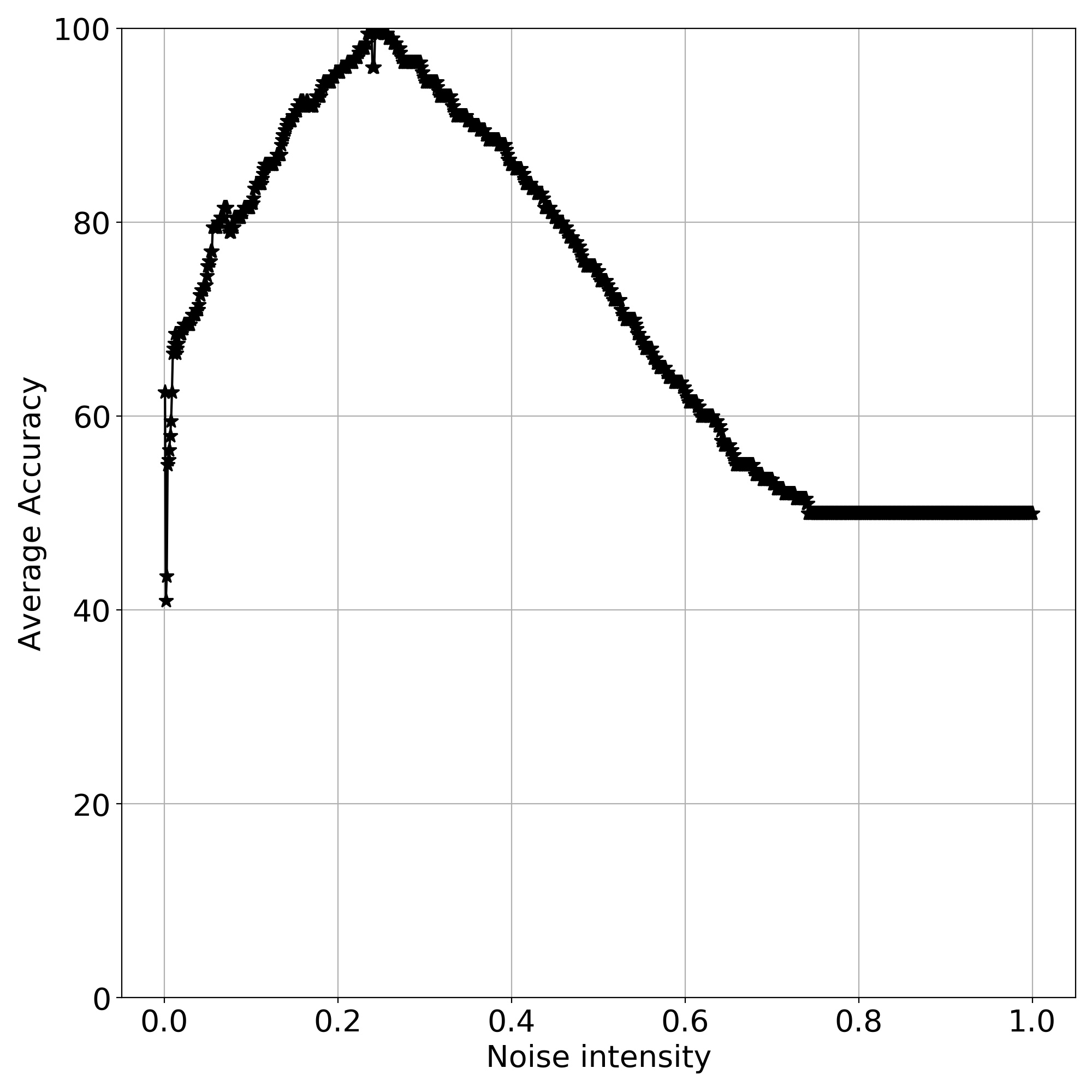}
		\caption{}\label{Fig_SR_plot_chaosnet_single_variable_classification}
	\end{subfigure}
	
\caption{SR in ChaosNet with single GLS neuron: (\subref{Fig_single_variable_Classification}) Prey-Predator dataset for organism $X$.   (\subref{Fig_SR_plot_chaosnet_single_variable_classification}) Average Accuracy ($\%$) vs. noise intensity shows stochastic resonance.}

\end{figure}

Having described the \verb+ChaosNet+ NL architecture (Figure~\ref{FIg_model_archi}), we begin our study by exploring the functioning of a single GLS neuron in this architecture. In order to understand the SR effect in a single GLS neuron, we consider a hypothetical example of the feeding pattern of a cannibalistic species with a food chain depicted in Figure~\ref{Fig_PP}. Specifically, we consider the problem of survival of a single organism $X$ of this species whose size is 0.5 units. All those organisms whose size is $\leq 0.5$ are a potential prey whereas those whose size is $>0.5$ is a potential predator. For survival, this specific organism $X$ needs to solve a binary classification problem. To this end, we assume that $X$ has a single GLS neuron that fires upon receiving a stimulus from the environment. The stimulus, say light reflected off the approaching organism, encodes the size of that organism, and we assume that it could be any value in the range $[0,1]$. The problem now reduces to determining whether the stimulus corresponds to the label `prey' (Class-0) or label `predator' (Class-1). An example of several stimuli received by the organism $X$ is depicted in Figure~\ref{Fig_single_variable_Classification} where Class-0 (Prey) and Class-1 (Predator) are marked with black stars and red circles respectively.

In order to solve the binary classification problem for survival, organism $X$ employs the \verb+ChaosNet+ NL algorithm~\cite{balakrishnan2019chaosnet}. Briefly, this is accomplished as follows. The single GLS neuron of organism $X$ fires chaotically from an initial neural activity (of $q$ units) until it matches the value of the stimulus. This chaotic neuronal trace encodes all the necessary information which the organism $X$ needs to determine whether the stimulus is Class-0 or Class-1. Particularly, the following four features are used -- firing time, firing rate, energy and entropy of neural trace. In the training phase, where the labels are assumed to be known, the organism $X$ learns the mean representation vector (the aforementioned 4 features) for each class (80 instances) and in the testing phase (20 instances per class), the organism $X$ estimates the label. This $(80,20)$ is the standard training-test distribution that we find in machine learning applications. From an evolutionary point of view, the interesting question to ask is - {\it how can natural selection optimize the biological aspects of organism $X$ in order for it to have the highest possible survival rate where it is able to correctly classify the stimulus as predator or prey?} Assuming that the organism can't afford to have more than one internal neuron (each neuron comes with a biological cost), the only available biological features are -- A) the type of GLS neuron and B) initial neural activity (this corresponds to memory) of this single internal neuron. The type of GLS neuron is determined by the value of the parameter $b$ (we assume the skew-tent map which is a type of 1D GLS map\footnote{In reality, it is possible for nature to evolve different types of neurons - which is very much the case with the brain. But for simplicity, we restrict to skew-tent map in this example.}) and the initial neural activity is represented by $q$. Both $b$ and $q$ in our example can take values in the range $(0,1)$. In our discussion so far, we have omitted a very important factor namely the {\it noise} that is inherent in the environment. Noise, as we very well know, is unavoidable and the stimulus (the light reflected off from the approaching organism) is inevitably distorted by ambient noise. In our efforts to optimize $b$ and $q$ to give the highest survival rate for organism $X$, we are also interested in asking the question -- {\it is the presence of ambient noise always detrimental to the decision making process of the GLS neuron, irrespective of the level of noise?} In other words, one would naively expect that even the presence of a tiny amount of noise can only degrade the performance of the chaotic GLS neuron thereby reducing the survival rate of organism $X$ (not to speak of high levels of noise where the performance is expected to much worse).

To answer the above two questions -- one pertaining to the best possible values of $q$ and $b$ that maximizes survival rate and the other to the role of the level of noise on the performance of GLS neuron for decision making -- we need to solve an optimization problem. To this end, we performed a five fold cross-validation on the dataset in Figure~\ref{Fig_single_variable_Classification} with $q = 0.25$, $b = 0.96$ and noise intensity varying from $0.001$ to $1$ in steps of $0.001$. For each trial of the 5-fold validation, we used $100$ data instances per class with a $(80,20)$ training-test split. The \verb+ChaosNet+ algorithm is run with this single GLS neuron (refer to~\cite{balakrishnan2019chaosnet, harikrishnan2020neurochaos} for further details). We report an average accuracy of $100 \%$ for noise intensities ranging from $0.248 $ to $0.253$. Figure~\ref{Fig_SR_plot_chaosnet_single_variable_classification} represents the average accuracy vs. noise intensity. 

From Figure~\ref{Fig_SR_plot_chaosnet_single_variable_classification} we are able to see that the relationship of noise to survival rate (average accuracy) is not monotonic, as we had originally expected. Rather, we see the phenomenon of Stochastic Resonance exhibited by Neurochaos Learning architecture. In other words, ambient noise is not always detrimental to the performance of the GLS neuron for the survival of the organism $X$. An intermediate amount of noise, in fact, maximizes the survival rate of organism $X$. Our hypothetical example is not very far from reality. In the case of feeding behaviour of paddle fish~\cite{russell1999paddle_fish} and in the physiological experiments on crayfish mechanorecetors, stochastic resonance has been reported. We conjecture that it may very well be the case of noise interacting with neurochaos leading to stochastic resonance in these real-world biological examples as well. 

We found that SR is exhibited for several other settings of $q$ and $b$ as well, but are not reported here for lack of space. The particular choice of $q=0.25$ and $b=0.96$ is motivated by the fact that SR enables a $100 \%$ accuracy for these settings (with an intermediate level of noise as seen in Figure~\ref{Fig_SR_plot_chaosnet_single_variable_classification}). However, such a performance is not unique to this particular choice of $q$ and $b$.

\subsection{SR in Signal Detection using ChaosNet NL~\label{SubSec:Signal_Detection}}   
Traditionally SR is shown to be exhibited in bistable system where there is an element of periodic forcing and stochastic component~\cite{bruno2000stochastic}. This notion is used to detect signals which are sub-threshold. A combination of a sub-threshold periodic and noisy signal provides a peak in detection of the frequency of the periodic signal. In this section, we demonstrate SR in signal detection using ChaosNet NL architecture. 

There are different ways to quantify SR for signal detection.  Some of the commonly used techniques are cross-correlation coefficient, signal to noise ratio, and mutual information~\cite{das2004quantifying_SR}. In this work, we use cross correlation coefficient as a measure to evaluate the SR behaviour in \verb+ChaosNet+ for signal detection.

\begin{figure}[h!]
\centering
	\begin{subfigure}{0.35\linewidth}
		\centering
		\includegraphics[width=1\linewidth]{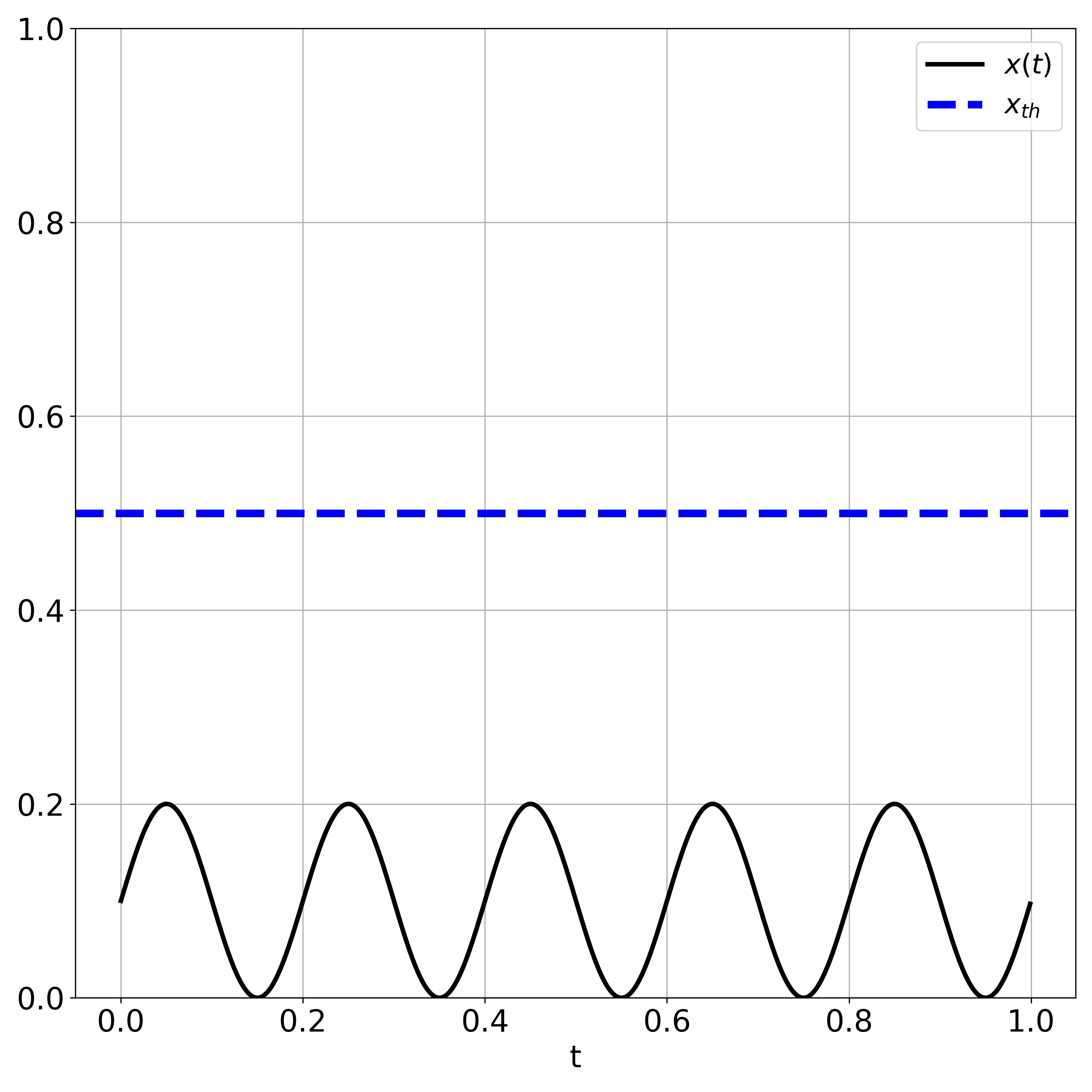}
		\caption{}\label{Fig_subthreshold_signal}
	\end{subfigure}
	\begin{subfigure}{0.35\linewidth}
		\centering
		\includegraphics[width=1\linewidth]{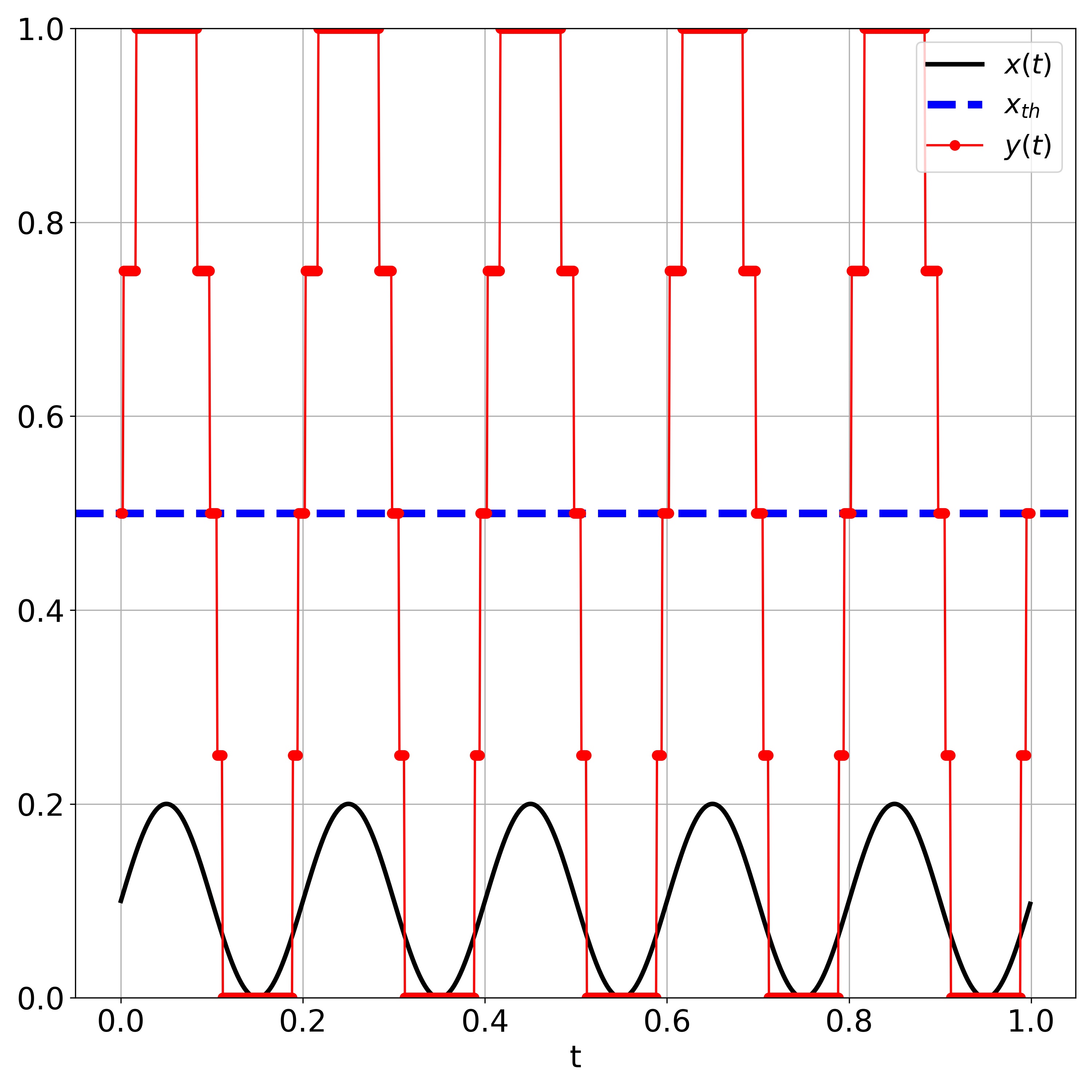}
		\caption{}\label{Fig_output_signal_detection}
	\end{subfigure}

	\begin{subfigure}{0.35\linewidth}
		\centering
		\includegraphics[width=1\linewidth]{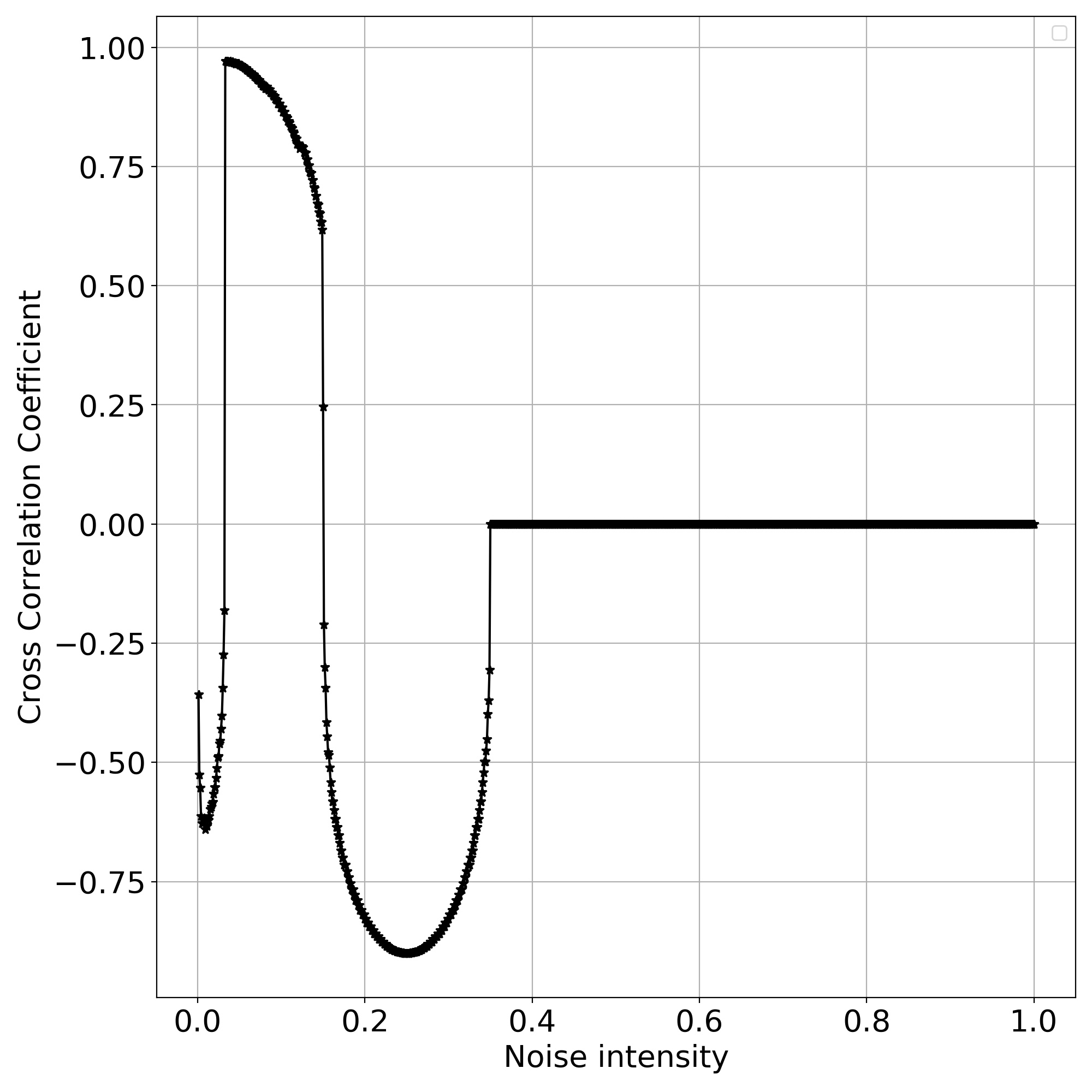}
		\caption{}\label{Fig_cross correlation coefficient_plot}
	\end{subfigure}
\caption{SR in signal detection using ChaosNet NL. (\subref{Fig_subthreshold_signal})  Sub-threshold signal $x(t)$ with threshold $x_{th} = 0.5$. The signal cannot be detected since it is below the threshold.  (\subref{Fig_output_signal_detection})  In the presence of intermediate amount of noise added to the input signal the single internal neuron in ChaosNet NL detects the weak amplitude signal and also preserves its frequency. (\subref{Fig_cross correlation coefficient_plot}) Cross correlation coefficient between $y(t)$ and $x(t)$  vs. noise intensity demonstrating SR. Whenever the variance of $y(t)$ was zero, we take $\rho = 0$.}
\end{figure}

\begin{figure}[h!]
\centering
	\begin{subfigure}{0.32\linewidth}
		\centering
		\includegraphics[width=1\linewidth]{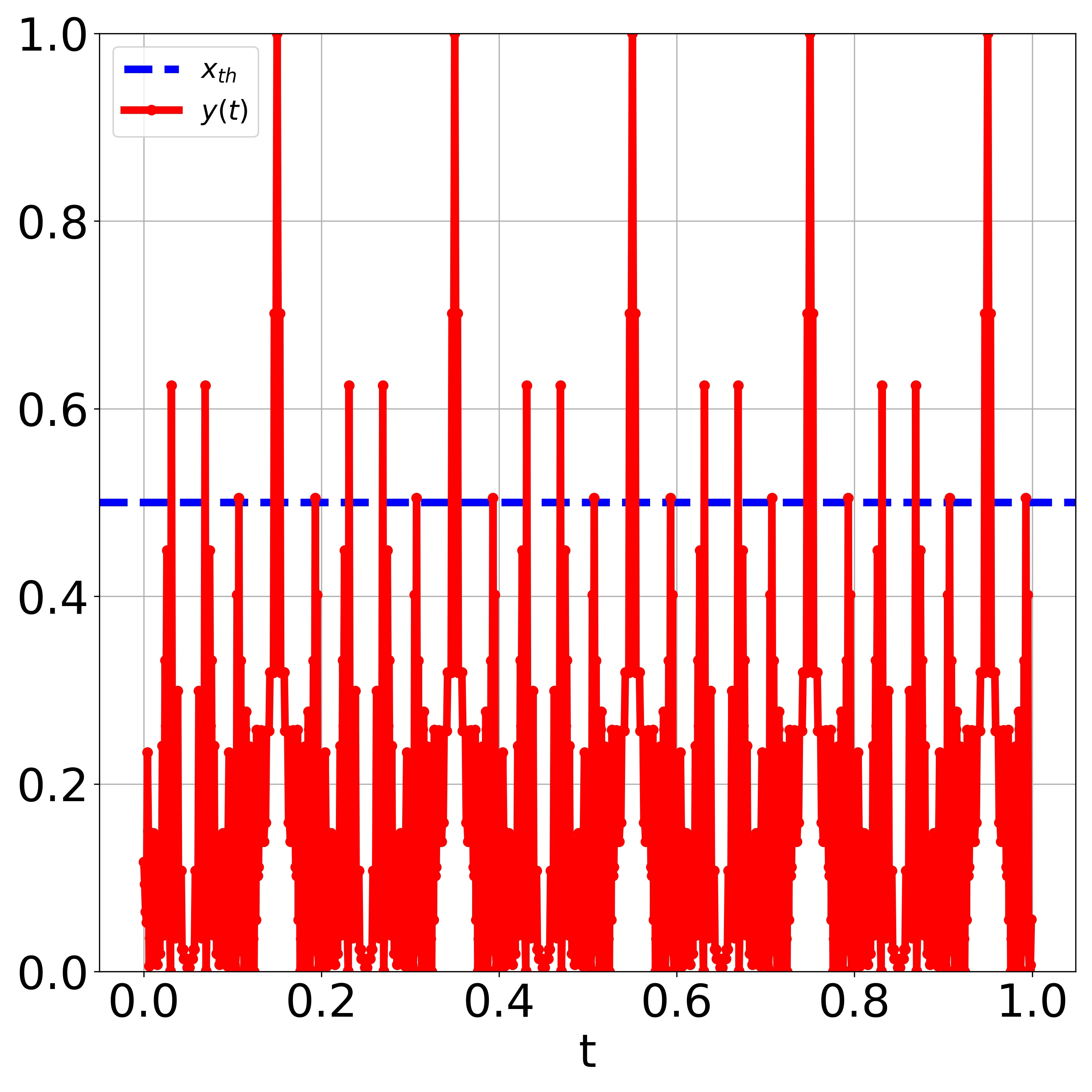}
		\caption{}\label{Fig_low_noise_sig_detection}
	\end{subfigure}
	\begin{subfigure}{0.32\linewidth}
		\centering
		\includegraphics[width=1\linewidth]{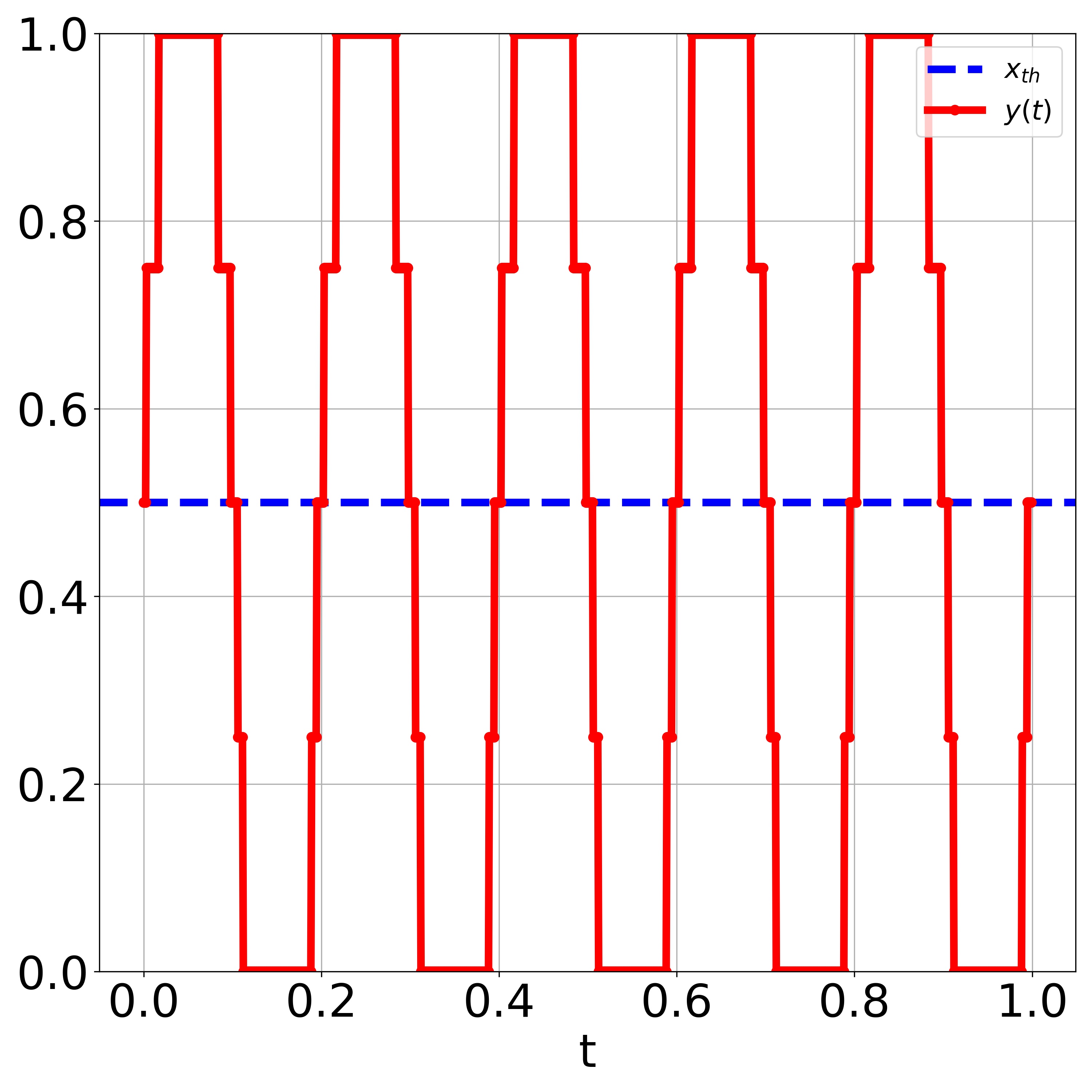}
		\caption{}\label{Fig_optimum_noise_sig_detection}
	\end{subfigure}
	\begin{subfigure}{0.32\linewidth}
		\centering
		\includegraphics[width=1\linewidth]{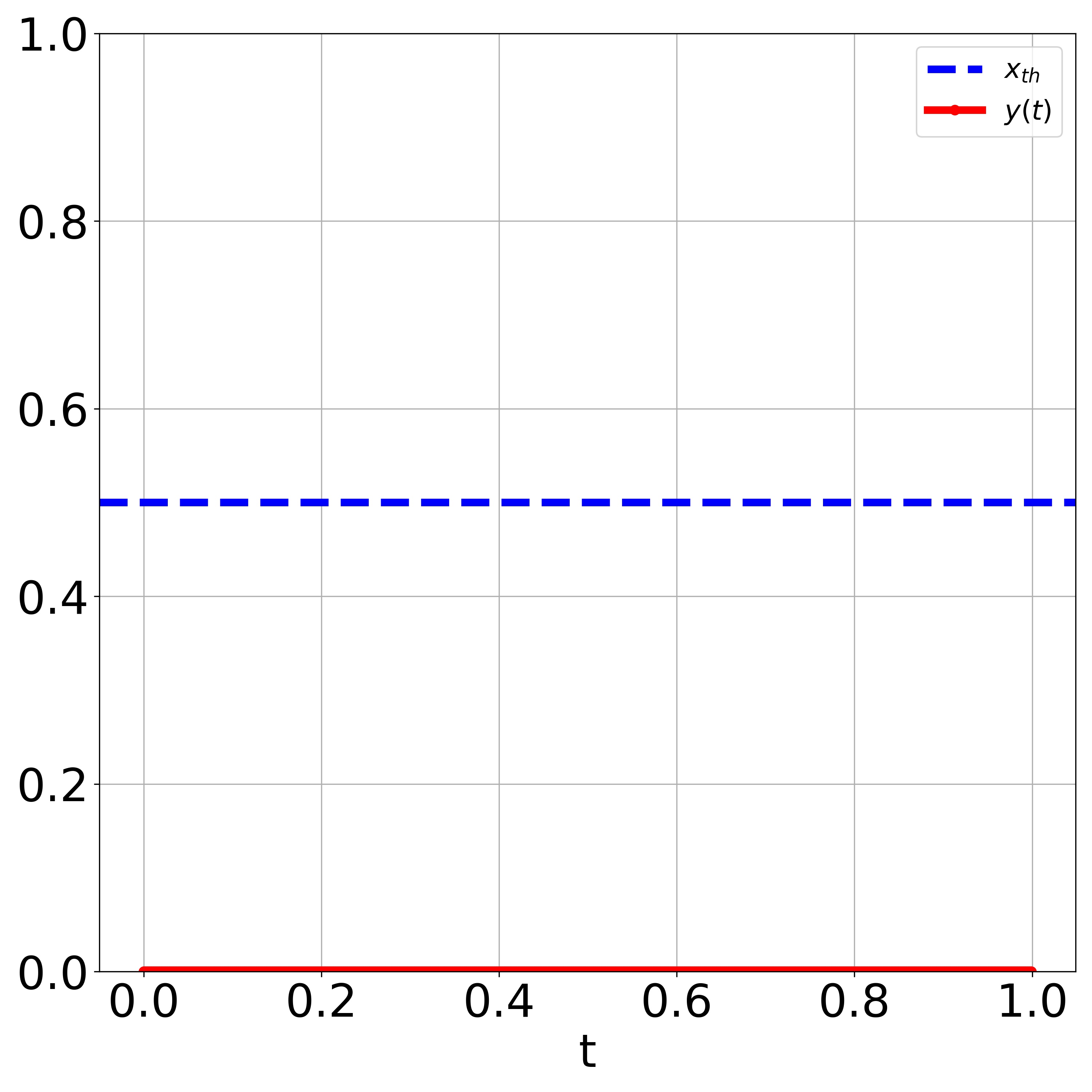}
		\caption{}\label{Fig_high_noise_sig_detection}
	\end{subfigure}

\caption{Intermediate noise is good for signal detection in NL. (\subref{Fig_low_noise_sig_detection}) With low noise ($\epsilon = 0.001$), signal detection is poor ($\rho = -0.357$). (\subref{Fig_optimum_noise_sig_detection}) With intermediate noise ($\epsilon = 0.033$), signal detection is good ($\rho = 0.971$). (\subref{Fig_high_noise_sig_detection}) With high noise ($\epsilon = 0.95$), signal detection is very poor ($\rho = 0$). Whenever the variance of $y(t)$ was zero, we take $\rho = 0$.}
\end{figure}

We consider a low amplitude information carrying periodic sinusoidal function $x(t) = \frac{A(\sin(2\pi5t)+1)}{2}$, where $A = 0.2$, and $0 \leq t \leq 1$s. Let the threshold be defined as $x_{th} = 0.5$. Clearly the low amplitude signal is below the threshold and hence the signal goes undetected (Figure~\ref{Fig_subthreshold_signal}). We pass $x(t) + \eta(t)$ as input to the \verb+ChaosNet+ architecture with a single GLS neuron with $q = 0.35$ and $b = 0.65$. Here, $\eta$ is noise which follows a uniform distribution with a range of $[-\epsilon +\epsilon]$, where $\epsilon$ is varied from $0.001$ to $1.0$ in steps of $0.001$. From the \verb+ChaosNet+ architecture we extract the normalized firing time\footnote{Normalization is done as follows: $y=\frac{Y-min(Y)}{max(Y)-min(Y)}$.}, $y(t)$,  shown in Figure~\ref{Fig_output_signal_detection} which closely tracks $x(t)$. We then compute the cross correlation coefficient $\rho$ between $y(t)$ and $x(t)$ for various noise intensities ($\epsilon$) as shown in Figure~\ref{Fig_cross correlation coefficient_plot}. We see that, for an intermediate amount of noise intensity, namely $\epsilon = 0.033$, we get a maximum cross correlation coefficient of $\rho = 0.971$ and for lower and higher noise intensities we get lower values of $\rho$ (this is because $y(t)$ fails to track $x(t)$ in these cases as demonstrated in Figure~\ref{Fig_low_noise_sig_detection}, ~\ref{Fig_high_noise_sig_detection}). This is the classic phenomenon of stochastic resonance. Thus, we have demonstrated SR in NL (\verb+ChaosNet+) with a single GLS neuron in both classification and signal detection scenarios. %

\section{SR in NL with multiple GLS Neurons\label{Sec:Exp_Evidence}}
We now move on to demonstrating SR in NL with more than one GLS neuron in both simulated and real-world datasets. 
\subsection{Simulated Data\label{SubSec:CCD}}
We consider a binary classification task of separating data instances belonging to two concentric circles which are either non-overlapping (CCD) or overlapping (OCCD).  The governing equations for OCCD are as follows:
\begin{eqnarray}
f_1 &=& r_i  \cos(\theta) + \alpha \eta, \label{eqn_ccd_1}  \\
 f_2 &=& r_i  \sin(\theta) + \alpha \eta, \label{eqn_ccd_2} 
\end{eqnarray}
where $i = \{0,1\}$ ($i = 0$ represents Class-0 and $i = 1$ represents Class-1), $r_0 = 0.6$, $r_1 = 0.4$, $\theta$ is varied from 0 to \ang{360}, $\alpha = 0.1$,  $\eta \sim  \mathcal{N}(\mu,\,\sigma)$, normal distribution, with $\mu = 0$ and 
$\sigma = 1$. For CCD, the value of $\alpha$ used in equation~\ref{eqn_ccd_1} and equation~\ref{eqn_ccd_2} is set to $0.01$. 

\begin{figure}[h!]
\centering
	\begin{subfigure}{0.45\linewidth}
		\centering
		\includegraphics[width=1\linewidth]{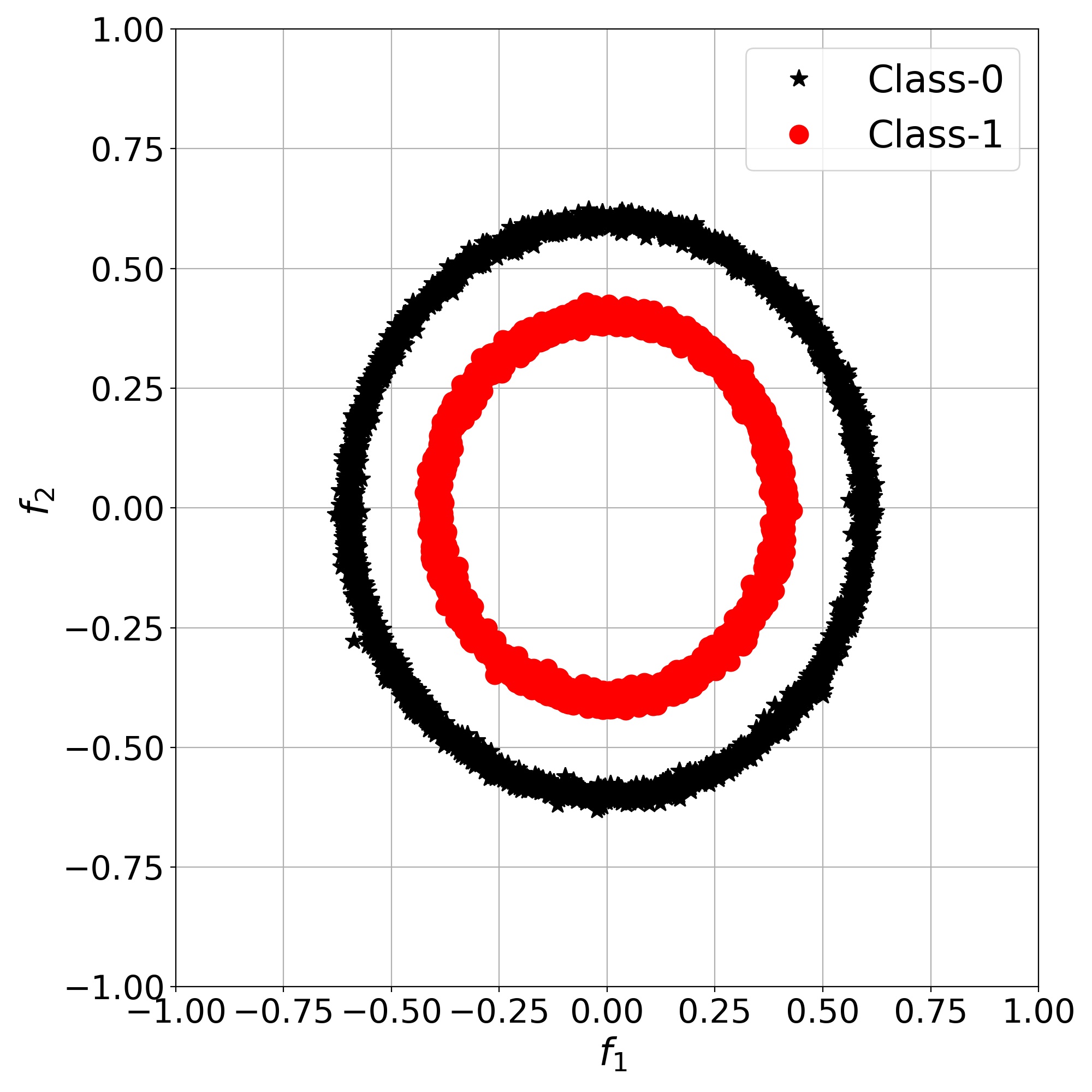}
		\caption{}\label{Fig_ccd_data}
	\end{subfigure}
	\begin{subfigure}{0.45\linewidth}
		\centering
		\includegraphics[width=1\linewidth]{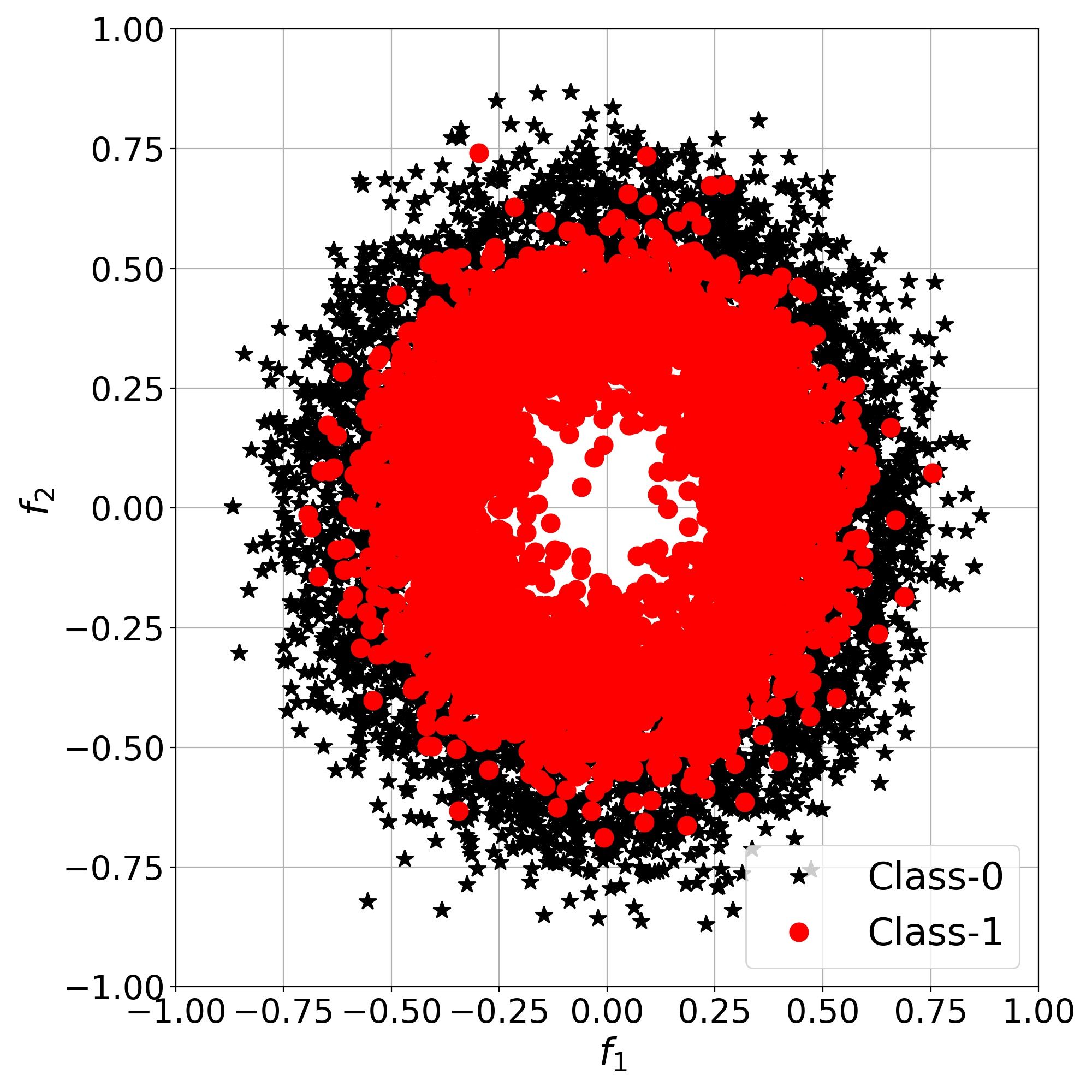}
		\caption{}\label{Fig_occd_data}
	\end{subfigure}
\caption{Simulated datasets for binary classification. (\subref{Fig_ccd_data}) Concentric Circle Data (CCD).   (\subref{Fig_occd_data}) Overlapping Concentric Circle (OCCD).}

	\label{fig:3:2}
\end{figure}

Figure~\ref{Fig_ccd_data} and Figure~\ref{Fig_occd_data} depict the non-overlapping (CCD) and overlapping (OCCD) data respectively. The dataset details for CCD and OCCD experiments are provided in Table~\ref{Table_CCD_OCCD}.

\begin{table}[!h]
\centering
\caption{Dataset details of synthetically generated Concentric Circle Data (CCD) and Overlapping Concentric Circle Data (OCCD).}
\begin{tabular}{|c|c|c|}
\hline
Dataset & \begin{tabular}[c]{@{}c@{}c@{}}CCD \\ \end{tabular} & OCCD \\ \hline
\# Classes & 2 & 2 \\ \hline
\begin{tabular}[c]{@{}c@{}}\# Training \\ instances per class\end{tabular} & (2513, 2527) & (2513, 2527) \\ \hline
\begin{tabular}[c]{@{}c@{}}\# Testing\\ instances per class\end{tabular} & (1087, 1073) & (1087, 1073) \\ \hline
\end{tabular}
\label{Table_CCD_OCCD}
\end{table}

\subsubsection{SR in ChaosNet NL on CCD and OCCD}
\begin{figure}[h!]
\centering
	\begin{subfigure}{0.35\linewidth}
		\centering
		\includegraphics[width=1\linewidth]{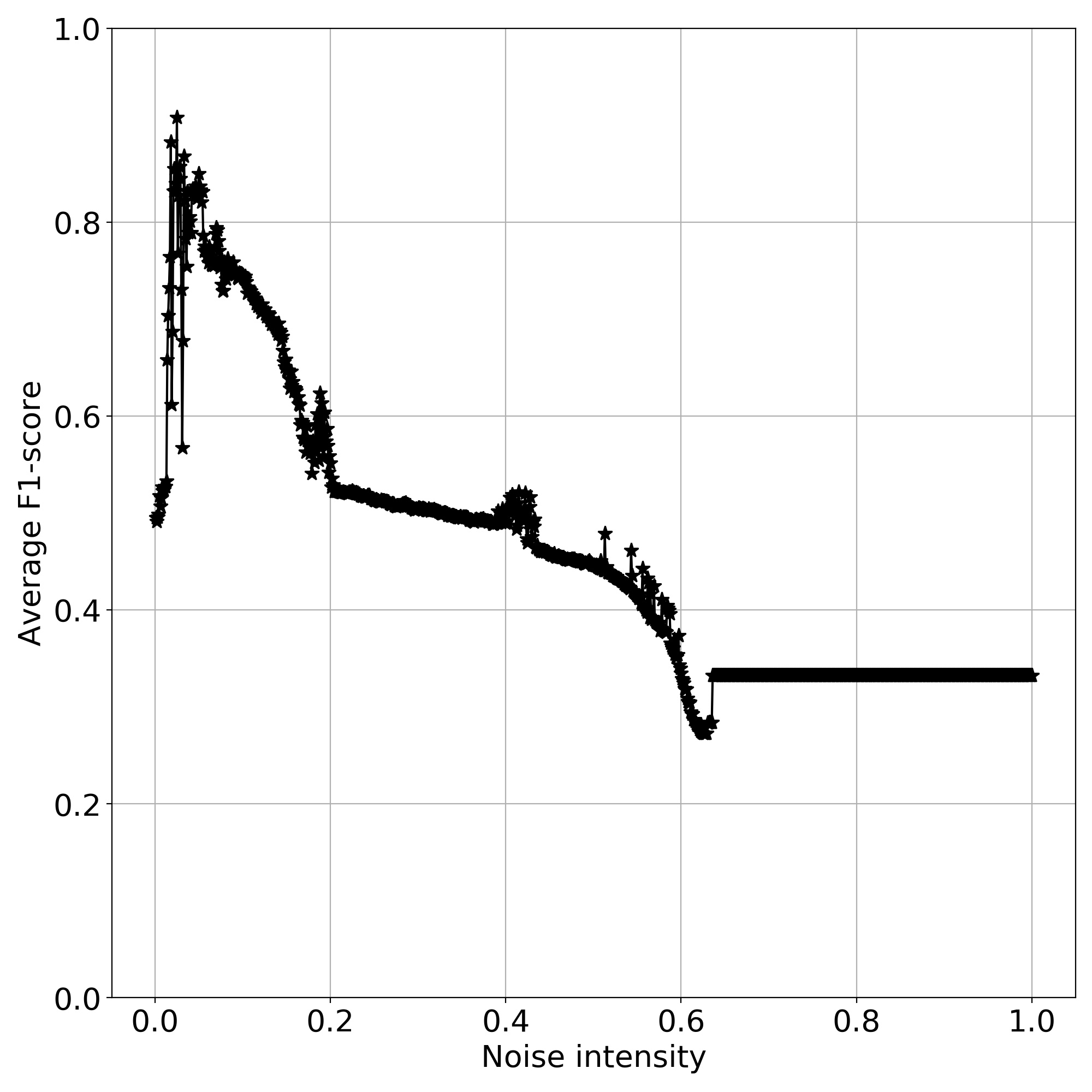}
		\caption{}\label{Fig_chaosnet_ccd_noise_intensity}
	\end{subfigure}
	\begin{subfigure}{0.35\linewidth}
		\centering
		\includegraphics[width=1\linewidth]{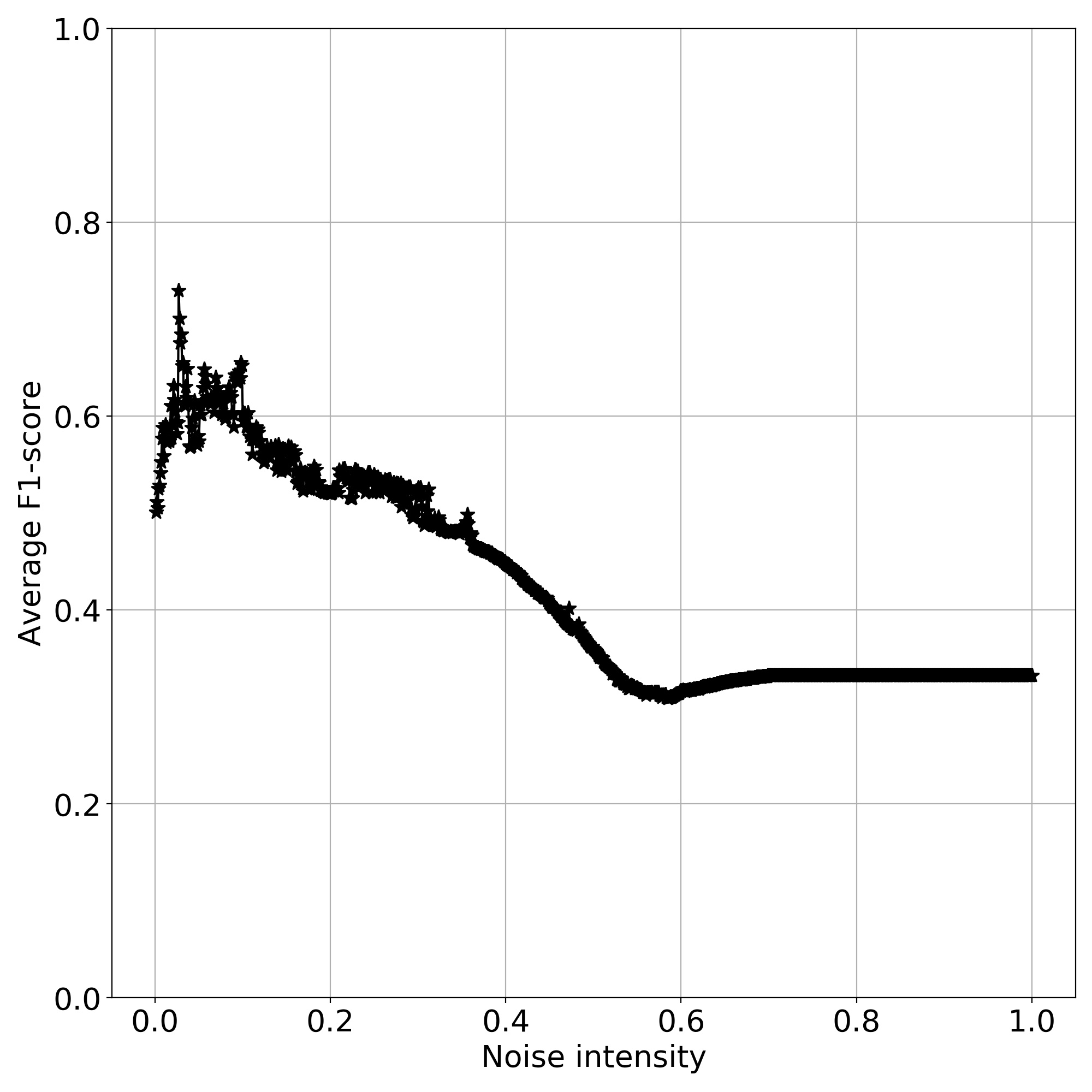}
		\caption{}\label{Fig_chaosnet_occd_noise_intensity}
	\end{subfigure}
	
\caption{SR in ChaosNet NL. (\subref{Fig_chaosnet_ccd_noise_intensity})  Average macro F1-score vs. Noise intensity for Concentric Circle Data (CCD).   (\subref{Fig_chaosnet_occd_noise_intensity}) Average macro F1-score vs. Noise intensity for Overlapping Concentric Circle (OCCD). Both plots indicate local as well as global SR.}
	
\end{figure}

We first perform a five fold cross-validation to determine the appropriate noise intensities for the CCD and OCCD classification tasks. The noise intensity ($\epsilon$) was varied from $0.001$ to $1$ with a step size of $0.001$. The initial neural activity and discrimination threshold were fixed to $q=0.21$ and $b=0.96$  respectively for CCD. In the case of OCCD, $q$ and $b$ are fixed to $0.23$ and $0.97$ respectively.
\begin{itemize}
    \item \textbf{CCD}: For noise intensity $\epsilon$ $= 0.025$, we get a maximum average F1-score $= 0.907$ in five fold cross-validation (Figure~\ref{Fig_chaosnet_ccd_noise_intensity}).
    \item \textbf{OCCD}: For noise intensity $\epsilon$ $= 0.027$, we get a maximum average F1-score $= 0.73$ in five fold cross-validation (Figure~\ref{Fig_chaosnet_occd_noise_intensity}).
\end{itemize}
Inspecting Figure~\ref{Fig_chaosnet_ccd_noise_intensity} and ~\ref{Fig_chaosnet_occd_noise_intensity}, we can make the following observations:
\begin{enumerate}
    \item Average F1-scores achieve global maxima for intermediate amount of noise intensities ($\epsilon$) in both cases (CCD and OCCD) indicating {\it global SR}.
    \item We observe several local maxima of average F1-scores in the above plots which we define as {\it local stochastic resonance} or {\it local SR}. 
    \item There could be multiple local SR in such tasks.
    \item It is also possible that multiple global SR exist for different settings of $b$, $q$ and $\epsilon$. 
    \item Such a rich behaviour of multiple SR (local and global) is due to the properties of chaos. 
    
\end{enumerate}

\subsection{SR in ChaosNet NL on real-world task}
In this section, we empirically show that SR is exhibited in \verb+ChaosNet+ NL  even for a real world classification task. The task is to identify spoken digits from a human speaker. Particularly, it is to classify the speech data into classes $0$ to $9$.

We considered the openly available Free Spoken Digit Dataset (FSDD)\footnote{\url{https://github.com/Jakobovski/free-spoken-digit-dataset}}. The dataset consists  of voice recordings of spoken digits from $0$ to $9$ of six speakers.  Out of the six speakers, for the purposes of this study, we considered the voice samples of only one speaker named `Jackson'. We have $50$ spoken digit recordings for each of the ten classes ($0$ to $9$)  sampled at $8$ $kHz$. The audio recordings are trimmed to remove the silence at the beginnings and ends. The maximum and minimum length of data instances are $7038$ and $2753$ respectively. In order to have the same length of data across all instances, we only considered the first $2753$ samples.  The  normalized Fourier coefficients of each data instance was extracted and input to \verb+ChaosNet+ NL. 

\begin{figure}[!h]
    \centerline{ \includegraphics[width=0.4\textwidth]{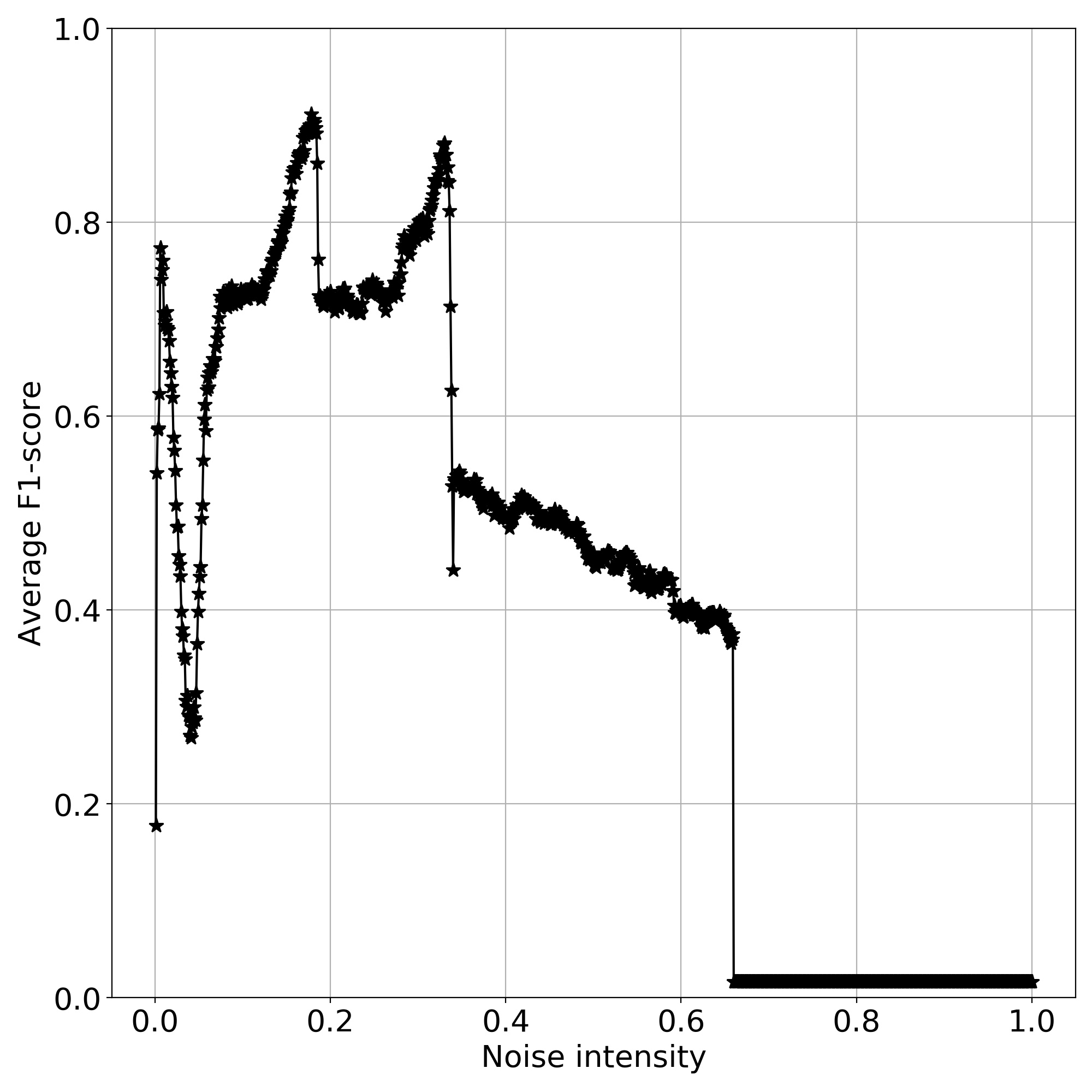}}
    
     \caption{SR in ChaosNet NL for spoken digit classification task. Average macro F1-score vs. Noise intensity.}
    \label{Fig_Speech_data_jackson_chaosnet}
    \end{figure}

We did a five fold cross-validation using $400$ data instances ($40$ data instances for each class) to determine the optimum noise intensity that yields the best performance. The noise intensity was varied from $0.001$ to $1.0$ in steps of $0.001$. For $q = 0.34$, $b = 0.499$ and a noise intensity $= 0.178$, we get a maximum macro average F1-score $= 0.911$. Figure~\ref{Fig_Speech_data_jackson_chaosnet} depicts the performance of \verb+ChaosNet+ NL with varying noise intensities and once again the familiar SR is seen (with local and global SR as defined in the previous section).
\subsection{Why is there SR in NL?}
Unlike traditional machine learning algorithms, stochastic resonance is not only found in neurochaos learning architecture (\verb+ChaosNet+) but also seen to contribute to a peak performance in classification tasks. This was demonstrated for NL with a single neuron as well as with multiple neurons, and for both simulated and real-world classification tasks. Even a simple sub-threshold signal detection task using NL exhibited strong SR.  

\begin{figure} [h!]
\centering
	\begin{subfigure}{0.32\linewidth}
		\centering
		\includegraphics[width=1\linewidth]{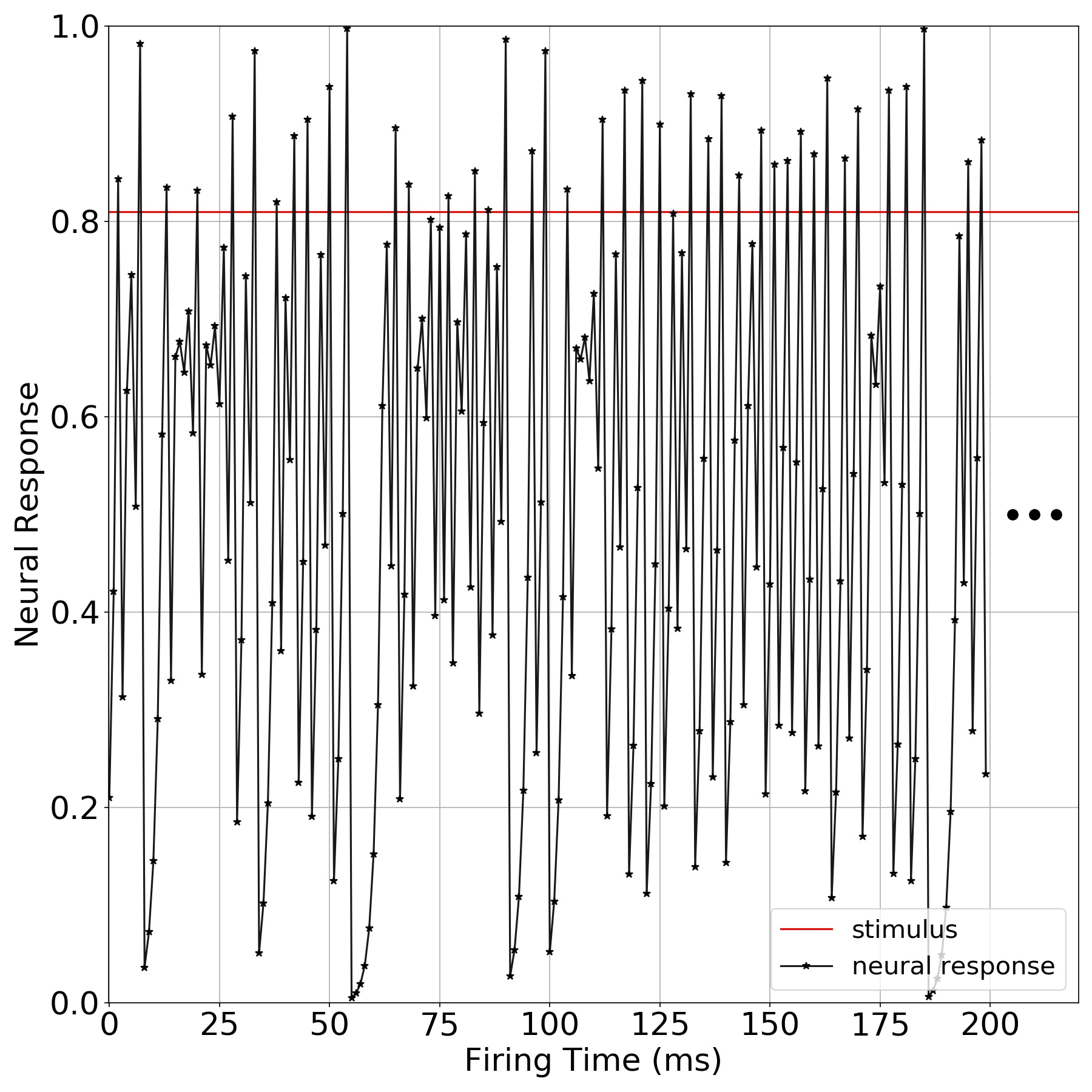}
		\caption{}\label{Fig_SR_Zero_Noise}
	\end{subfigure}
	\begin{subfigure}{0.32\linewidth}
		\centering
		\includegraphics[width=1\linewidth]{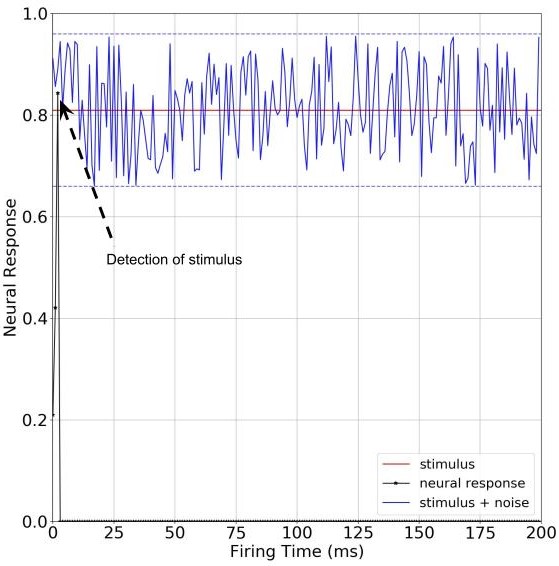}
		\caption{}\label{Fig_SR_High_Noise}
	\end{subfigure}
	\begin{subfigure}{0.32\linewidth}
		\centering
		\includegraphics[width=1\linewidth]{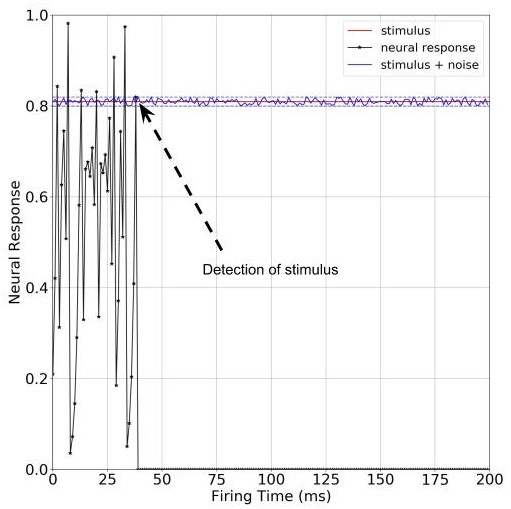}
		\caption{}\label{Fig_SR_Low_Noise}
	\end{subfigure}
\caption{Why SR in NL yields peak performance? (\subref{Fig_SR_Zero_Noise}) With zero noise, the neural trace is indefinite and hence firing time is undefined (infinite).  (\subref{Fig_SR_High_Noise}) With very high noise, the neural trace matches the stimulus in very short time yielding very low firing time. (\subref{Fig_SR_Low_Noise}) For medium level of noise, there is a sufficient length of neural trace which enables meaningful features to be extracted for learning.  This allows stimuli from different instances and different classes to have diverse length of neural traces with distinct features that enable peak classification performance in NL.}
	\label{fig:whySRinNL}
\end{figure}

The question that we now like to address is - {\it why is there SR in NL?} To answer this, we observe a single neural trace (or trajectory) of a single GLS neuron (of NL) corresponding to a single stimulus. Figure~\ref{fig:whySRinNL} shows the neural trace (in black) for three scenarios - (a) zero noise ($\epsilon = 0.0$), (b) high noise ($\epsilon = 0.15$) and (c) medium noise ($\epsilon = 0.01$). The target stimulus is indicated in red and the noisy stimulus (target $+$ noise) in blue. The GLS neuron stops firing only when the neural activity matches the stimulus. The stopping time and the noise intensity is inversely proportional. For zero noise, the GLS neuron fires indefinitely without stopping since the probability of the neural activity becoming equal to the stimulus is zero. For a very high noise intensity ($epsilon \approx 1$), the stopping time is very low (could be even zero) since the huge variation in the noisy stimulus ensures that the neural activity already matches the stimulus. It is only for a medium level of noise that there is a sufficient length of neural trace which enables meaningful features to be extracted for learning.  This allows stimuli from different instances and different classes to have diverse length of neural traces with distinct features that enable efficient learning for classification.  From a biological point of view, neurons in the brain can neither fire indefinitely nor not fire at all. It seems intuitive to posit that a rich pattern of firing is necessary for learning in the human brain. 
\section{Conclusions and Future Work\label{Sec:Conclusion}}
Noise is always contextual. A universal mathematical definition of noise without contextual consideration does not exist. Noise cannot always be treated as an unwanted signal. Stochastic Resonance is one such counter intuitive phenomenon where the constructive role of noise is seen to contribute a performance boost in certain non-linear systems. In this study, we highlight for the first time how stochastic resonance is naturally manifesting in \verb+ChaosNet+ neurochaos learning architecture for classification and signal detection. Future work involves considering NL with spiking neuronal models instead of GLS and exploring whether SR has a similar role in performance. 

Our study paves the way for potentially unravelling how learning happens in the human brain. There is ample amount of empirical evidence to suggest that chaos is inherent at the neuronal level in the brain, as well as at different spatiotemporal scales~\cite{therechaos1, therechaos2}. Also, given the enormous complexity of brain networks, neuronal noise and interference are unavoidable~\cite{czanner2015measuring}. In spite of these challenges, the brain does a phenomenal job in learning - currently unmatched by ANNs if power consumption is factored (the brain is known to operate at $\approx 12.6$ watts)~\cite{jabr_2012_scientific_american}. 

Thus, noise has a definite role to play in cognitive systems that can be modeled as non-linear chaotic learning systems (both NL architectures like \verb+ChaosNet+ and the human brain). Noise optimizes the selection of diverse chaotic neural traces with distinct features that enables efficient learning. SR seems to be not only inevitable but indispensable in cognitive systems. 

The code used for the study of SR in NL is available here: \url{https://github.com/HarikrishnanNB/stochastic_resonance_and_nl}.
\section*{Acknowledgment}
Harikrishnan N. B. thanks ``The University of Trans-Disciplinary Health Sciences and Technology (TDU)'' for permitting this research as part of the PhD programme. The authors gratefully acknowledge the financial support of Tata Trusts. We dedicate this work to the founder and Chancellor of Amrita Vishwa Vidyapeetham-Sri Mata Amritanandamayi Devi (AMMA) who continuously inspires us by her dedication and commitment in serving humanity with love and compassion.
%
%
%
%

\begin{thebibliography}{10}

\bibitem{hassabis2017neuroscience_AI}
Demis Hassabis, Dharshan Kumaran, Christopher Summerfield, and Matthew
  Botvinick.
\newblock Neuroscience-inspired artificial intelligence.
\newblock {\em Neuron}, 95(2):245--258, 2017.

\bibitem{azevedo2009equal_neurons_in_brain}
Frederico~AC Azevedo, Ludmila~RB Carvalho, Lea~T Grinberg, Jos{\'e}~Marcelo
  Farfel, Renata~EL Ferretti, Renata~EP Leite, Wilson~Jacob Filho, Roberto
  Lent, and Suzana Herculano-Houzel.
\newblock Equal numbers of neuronal and nonneuronal cells make the human brain
  an isometrically scaled-up primate brain.
\newblock {\em Journal of Comparative Neurology}, 513(5):532--541, 2009.

\bibitem{therechaos2}
Henri Korn and Philippe Faure.
\newblock Is there chaos in the brain? ii. experimental evidence and related
  models.
\newblock {\em Comptes rendus biologies}, 326(9):787--840, 2003.

\bibitem{faisal2008noise_nervous_system}
A~Aldo Faisal, Luc~PJ Selen, and Daniel~M Wolpert.
\newblock Noise in the nervous system.
\newblock {\em Nature reviews neuroscience}, 9(4):292--303, 2008.

\bibitem{brown1828brief}
Robert Brown F.R.S. Hon. M.R.S.E. \& R.I. Acad. V.P.L.S.
\newblock XXVII. a brief account of microscopical observations made in the
  months of june, july and august 1827, on the particles contained in the
  pollen of plants; and on the general existence of active molecules in organic
  and inorganic bodies.
\newblock {\em The Philosophical Magazine}, 4(21):161--173, 1828.

\bibitem{einstein1905brownian}
Albert Einstein.
\newblock {\"U}ber die von der molekularkinetischen theorie der w{\"a}rme
  geforderte bewegung von in ruhenden fl{\"u}ssigkeiten suspendierten teilchen.
\newblock {\em Annalen der physik}, 4, 1905.

\bibitem{ising1926lgalvanometer}
Gustaf Ising.
\newblock LXXIII. a natural limit for the sensibility of galvanometers.
\newblock {\em The London, Edinburgh, and Dublin Philosophical Magazine and
  Journal of Science}, 1(4):827--834, 1926.

\bibitem{nyquist1928thermal}
Harry Nyquist.
\newblock Thermal agitation of electric charge in conductors.
\newblock {\em Physical review}, 32(1):110, 1928.

\bibitem{johnson1928thermal}
John~Bertrand Johnson.
\newblock Thermal agitation of electricity in conductors.
\newblock {\em Physical review}, 32(1):97, 1928.

\bibitem{shannon1948mathematical}
Claude~E Shannon.
\newblock A mathematical theory of communication.
\newblock {\em The Bell system technical journal}, 27(3):379--423, 1948.

\bibitem{berger2012climatic}
Andr{\'e}~L Berger.
\newblock {\em Climatic Variations and Variability: Facts and Theories: NATO
  Advanced Study Institute First Course of the International School of
  Climatology, Ettore Majorana Center for Scientific Culture, Erice, Italy,
  March 9--21, 1980}, volume~72.
\newblock Springer Science \& Business Media, 2012.

\bibitem{benzi1982_climate_SR}
Roberto Benzi, Giorgio Parisi, Alfonso Sutera, and Angelo Vulpiani.
\newblock Stochastic resonance in climatic change.
\newblock {\em Tellus}, 34(1):10--16, 1982.

\bibitem{fauve1983SR_electronic_circuit}
St{\'e}phan Fauve and F~Heslot.
\newblock Stochastic resonance in a bistable system.
\newblock {\em Physics Letters A}, 97(1-2):5--7, 1983.

\bibitem{bulsara1991_SR_neuron}
Adi Bulsara, EW~Jacobs, Ting Zhou, Frank Moss, and Laszlo Kiss.
\newblock Stochastic resonance in a single neuron model: Theory and analog
  simulation.
\newblock {\em Journal of Theoretical Biology}, 152(4):531--555, 1991.

\bibitem{leonard1994_SR_chemical_reaction}
David~S Leonard and LE~Reichl.
\newblock Stochastic resonance in a chemical reaction.
\newblock {\em Physical Review E}, 49(2):1734, 1994.

\bibitem{mcdonnell2008stochastic}
Mark~D McDonnell, Nigel~G Stocks, Charles~EM Pearce, and Derek Abbott.
\newblock Stochastic resonance.
\newblock {\em Stochastic Resonance}, 2008.

\bibitem{douglass1993_SR_cray_fish}
John~K Douglass, Lon Wilkens, Eleni Pantazelou, and Frank Moss.
\newblock Noise enhancement of information transfer in crayfish
  mechanoreceptors by stochastic resonance.
\newblock {\em Nature}, 365(6444):337--340, 1993.

\bibitem{longtin1993_SR_neuron_model}
Andr{\'e} Longtin.
\newblock Stochastic resonance in neuron models.
\newblock {\em Journal of statistical physics}, 70(1-2):309--327, 1993.

\bibitem{russell1999paddle_fish}
David~F Russell, Lon~A Wilkens, and Frank Moss.
\newblock Use of behavioural stochastic resonance by paddle fish for feeding.
\newblock {\em Nature}, 402(6759):291--294, 1999.

\bibitem{chatterjee2001noise_cochlear_implant}
Monita Chatterjee and Mark~E Robert.
\newblock Noise enhances modulation sensitivity in cochlear implant listeners:
  Stochastic resonance in a prosthetic sensory system?
\newblock {\em Journal of the Association for Research in Otolaryngology},
  2(2):159--171, 2001.

\bibitem{bulsara2010logical}
Adi~R Bulsara, Anna Dari, William~L Ditto, K~Murali, and Sudeshna Sinha.
\newblock Logical stochastic resonance.
\newblock {\em Chemical Physics}, 375(2-3):424--434, 2010.

\bibitem{ikemoto2018noise_SR_NN}
Shuhei Ikemoto, Fabio DallaLibera, and Koh Hosoda.
\newblock Noise-modulated neural networks as an application of stochastic
  resonance.
\newblock {\em Neurocomputing}, 277:29--37, 2018.

\bibitem{schilling2020intrinsic_SR_Speech}
Achim Schilling, Richard Gerum, Alexandra Zankl, Holger Schulze, Claus Metzner,
  and Patrick Krauss.
\newblock Intrinsic noise improves speech recognition in a computational model
  of the auditory pathway.
\newblock {\em bioRxiv}, 2020.

\bibitem{harikrishnan2020neurochaos}
NB~Harikrishnan and Nithin Nagaraj.
\newblock Neurochaos inspired hybrid machine learning architecture for
  classification.
\newblock In {\em 2020 International Conference on Signal Processing and
  Communications (SPCOM)}, pages 1--5. IEEE, 2020.

\bibitem{balakrishnan2019chaosnet}
Harikrishnan~Nellippallil Balakrishnan, Aditi Kathpalia, Snehanshu Saha, and
  Nithin Nagaraj.
\newblock Chaosnet: A chaos based artificial neural network architecture for
  classification.
\newblock {\em Chaos: An Interdisciplinary Journal of Nonlinear Science},
  29(11):113125, 2019.

\bibitem{therechaos1}
Philippe Faure and Henri Korn.
\newblock Is there chaos in the brain? i. concepts of nonlinear dynamics and
  methods of investigation.
\newblock {\em Comptes Rendus de l'Acad{\'e}mie des Sciences-Series
  III-Sciences de la Vie}, 324(9):773--793, 2001.

\bibitem{laleh2020chaotic}
Touraj Laleh, Mojtaba Faramarzi, Irina Rish, and Sarath Chandar.
\newblock Chaotic continual learning.
\newblock {\em 4-th Lifelong Learning Workshop at ICML}, 2020.

\bibitem{bruno2000stochastic}
Bruno And{\`o} and Salvatore Graziani.
\newblock {\em Stochastic resonance: theory and applications}.
\newblock Springer Science \& Business Media, 2000.

\bibitem{das2004quantifying_SR}
Aruneema Das, NG~Stocks, A~Nikitin, and EL~Hines.
\newblock Quantifying stochastic resonance in a single threshold detector for
  random aperiodic signals.
\newblock {\em Fluctuation and Noise Letters}, 4(02):L247--L265, 2004.

\bibitem{czanner2015measuring}
Gabriela Czanner, Sridevi~V Sarma, Demba Ba, Uri~T Eden, Wei Wu, Emad Eskandar,
  Hubert~H Lim, Simona Temereanca, Wendy~A Suzuki, and Emery~N Brown.
\newblock Measuring the signal-to-noise ratio of a neuron.
\newblock {\em Proceedings of the National Academy of Sciences},
  112(23):7141--7146, 2015.

\bibitem{jabr_2012_scientific_american}
Ferris Jabr.
\newblock Does thinking really hard burn more calories?
\newblock {\em Scientific American}, Jul 2012.

\end{thebibliography}

\end{document}